\documentclass{kapproc} 

\RequirePackage{graphicx}%
\RequirePackage{epsf}%
\input{psfig}

\upperandlowercase
\let\footnote\savefootnote
\let\footnotetext\savefootnotetext 
 
\kluwerbib 

\begin{document}


\articletitle{The Starburst IMF -- An Impossible Measurement?}


\chaptitlerunninghead{Starburst IMF}



 \author{Bernhard R. Brandl\altaffilmark{1} \& 
         Morten Andersen\altaffilmark{2}}
 \affil{\altaffilmark{1}Leiden Observatory, P.O. Box 9513, 
                        2300 RA Leiden, The Netherlands\\
        \altaffilmark{2}Steward Observatory, 933 N Cherry Ave., 
	                Tucson AZ 85721, USA}
 \email{brandl@strw.leidenuniv.nl}





\begin{abstract}
The starburst IMF is probably as much of theoretical interest and
practical relevance as it is a subject of observational controversy.
In this conference paper we review the most common methods (star
counts, dynamical masses, and line ratios) to derive or constrain the
IMF, and discuss potential problems and shortcomings that often lead
to claims of anomalous IMF slopes and mass cut-offs.

\end{abstract}

\section{Introduction}
The IMF in regions of violent star formation activity, so-called
starbursts, is still a rather controversial issue.  For instance, the
ultra-luminous infrared galaxy Arp~220 with a total gas mass of
$1.6\times 10^{10} M_\odot$ Arp~220 has about five times the amount of
gas available in the Milky Way.  However, its star formation rate of
$300 M_\odot$/yr is about 200 times higher than in our Galaxy, and
-- in addition -- the regions of star formation are rather
concentrated near the center(s) of the galaxy.  

It may seem plausible that in such regions, where the physical
boundary conditions for star formation differ so substantially from
the solar neighborhood, the IMF may be different as well.  Such a
difference may be indicated by either a different IMF slope (``top
heavy''), a significantly higher low-mass cut-off (``$2-5 M_\odot$'')
or a different high-mass cut-off.

The observational verification of these differences is all but
trivial: there are few regions that can serve as starburst templates,
most of which are at large distances, have unknown morpholgies, and
are often heavily extincted.  For closer regions, which can be
resolved into individual stars, the dynamical range of typically more
than 10 magnitudes in flux differences between low and high mass stars
is a significant limitation.  Not surprisingly, the strongest support
for anomalous IMFs to date comes from the observations of distant
starburst galaxies or individual super star clusters.  While those
environments are indeed rather different from the local ones it should
also be kept in mind how little we know about these distant system as
the uncertainties in both observations and models are fairly large.
In this conference paper we discuss the general difficulties in the
IMF determination.

\section{Method I:  star counts}
The most common method, which has also be shown to work well in nearby
regions of quiescent star formation, is the direct detection of stars.
After photometric extraction of the sources and subtraction of
foreground and background field stars, the correction for
incompleteness (see below) takes care of instrumental limitations and
possibly of the extra contribution from unresolved binary systems.
Finally, the conversion from magnitudes to masses leads to a {\it
present-day} mass function, which is often, and not very precisely,
called the {\it initial} mass function as dynamical effects in young
clusters are (incorrectly) considered to be of little importance.

\begin{figure}[ht]
\centerline{\psfig{file=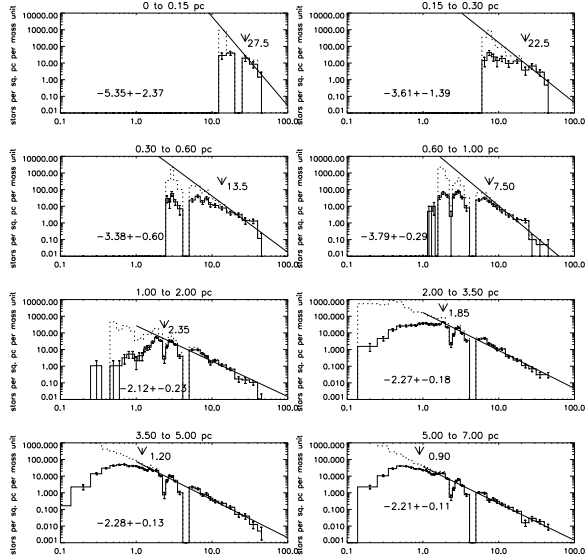,height=3in}}
\caption{The derived IMFs for R136 in radial bins using a \cite{sie00}
         3~Myr isochrone with half-solar metallicity, a distance
         modulus of 18.5 and an extinction of $A_V=2$ \cite{and04b}.}
\label{r136imf}
\end{figure}

The results from a recent study of the IMF in R136, the ionizing
cluster in the 30~Doradus nebula, are shown in Figure~\ref{r136imf} as
a function of radial distance \cite{and04b}.  The errors for the
innermost 1\,pc ($4''$) are very large and the derived slopes not very
meaningful.  However, outside the core region the slope agrees, within
the uncertainties, well with a \cite{sal55} slope and shows no
significant steepening with distance to the cluster center.

However, each step from the observed image to an IMF slope involves
potential difficulties, of which we can here address only a few:

\subsection{Problem: the high-mass stars}
At no time within a massive, coeval cluster's lifetime will it be
possible to observe all its members on the main sequence.  Many of the
most massive stars will have collapsed or exploded before the lower
mass stars reach the main sequence.  High mass stars show strong
luminosity evolution and mass loss during their short lifetimes.  In
particular the luminosity evolution of a given spectral subtype within
its lifetime can be larger than the initial difference between
different spectral subtypes.  It is therefore almost impossible to
accurately determine masses of O-stars from photometry alone (and
individual spectra are usually not available).  

There is also mounting evidence that the evolution of massive stars in
the core of a dense stellar cluster is significantly influenced by the
other cluster members.  \cite{por99} have shown in N-body simulations
that due to the high stellar densities mergers of massive stars occur
frequently, and by doing so increase its cross-section while the
cluster is becoming denser due to mass segregation, in some kind of
``runaway merger'' process.  In other words, a central massive star
grows steadily in mass through mergers with other stars in less than
$3-4$\,Myr, and the observed properties are not representative of the
initial mass distributions.

\subsection{Problem: the low-mass stars}
Besides the obvious difficulty of detecting relatively faint, low-mass
stars a significant uncertainty will be introduced by the fact that in
most young clusters they are still in their pre-main sequence (PMS)
phase.  (The fact that PMS stars are brighter than their main sequence
counterparts at infrared wavelengths, improves the situation
slightly).  However, the models for converting magnitudes to masses
depend critically on the cluster age which has typical uncertainties
in the order of $0.5~Myr$.  Usually all stars within a cluster are
assumed to be co-eval.  Furthermore, the conversion also depends on
the wavebands that were used and on the specific PMS model.  This
situation is illustrated in Table~\ref{tabpms}.
\begin{table}[ht]
\caption{Comparison of IMF slopes derived from the same data but with
         different PMS models, age assumptions, distance moduli or
         based on different filter bands.  The differences in the
         resulting slopes are striking!  Table from \cite{and04a}.}
\begin{tabular}{lcccc}
\hline
\it Model$^a$ & \it Age [Myr] & \it DM$^b$ & \it color & \it slope \cr
\hline
SDF & 1 & 13.9 & $J_s-K_s$ & $-2.20\pm 0.13$ \cr
SDF & 1 & 13.9 & $H-K_s$ & $-1.46\pm 0.23$ \cr
SDF & 1 & 14.3 & $J_s-K_s$ & $-2.41\pm 0.14$ \cr
SDF & 2 & 13.9 & $H-K_s$ & $-1.46\pm 0.34$ \cr
PS  & 1 & 13.9 & $J_s-K_s$ & $-1.70\pm 0.19$ \cr
\hline
\end{tabular}
\begin{tablenotes}
$^a$SDF -- \cite{sie00}; PS -- \cite{pal99}.
$^b$DM -- distance modulus
\end{tablenotes}
\label{tabpms}
\end{table}

The effects of heating, outflows and stellar winds \cite{lam95}, and
in particular the strong winds from O-stars \cite{chu97}, will cause
the residual gas to be removed from the cluster on rather short
timescales of $~1$\,Myr. If the gas cloud was originally virialized,
and the star formation efficiency is about 30\%, the cluster will
loose about 2/3 of its mass.  As a result, a signifcant fraction of
its young members will get lost, preferentially low-mass stars at the
high tail of the velocity dispersion and/or at large distances
\cite{kro02}.

This has important observational consequences: even for a rather
distant Galactic cluster like NGC\,3603 at 7\,kpc with an assumed
velocity dispersion of 5\,km/s the radial distances of initial cluster
stars can be as large as $100''$ after 1\,Myr (including a
$\cos(45^o)$ projection factor).  If the reference field taken to
estimate the field star population is closer than this radius there
are two problems: {\it (i)} stars that were born in the center are now
missing there, and {\it (ii)} those stars may now be mistaken as field
stars and subtracted from the cluster sample.  Both effects work in
the same direction and introduce an artificial flattening of the IMF.

\subsection{Problem: differential extinction}
\begin{figure}[ht]
\sidebyside
{\centerline{\psfig{file=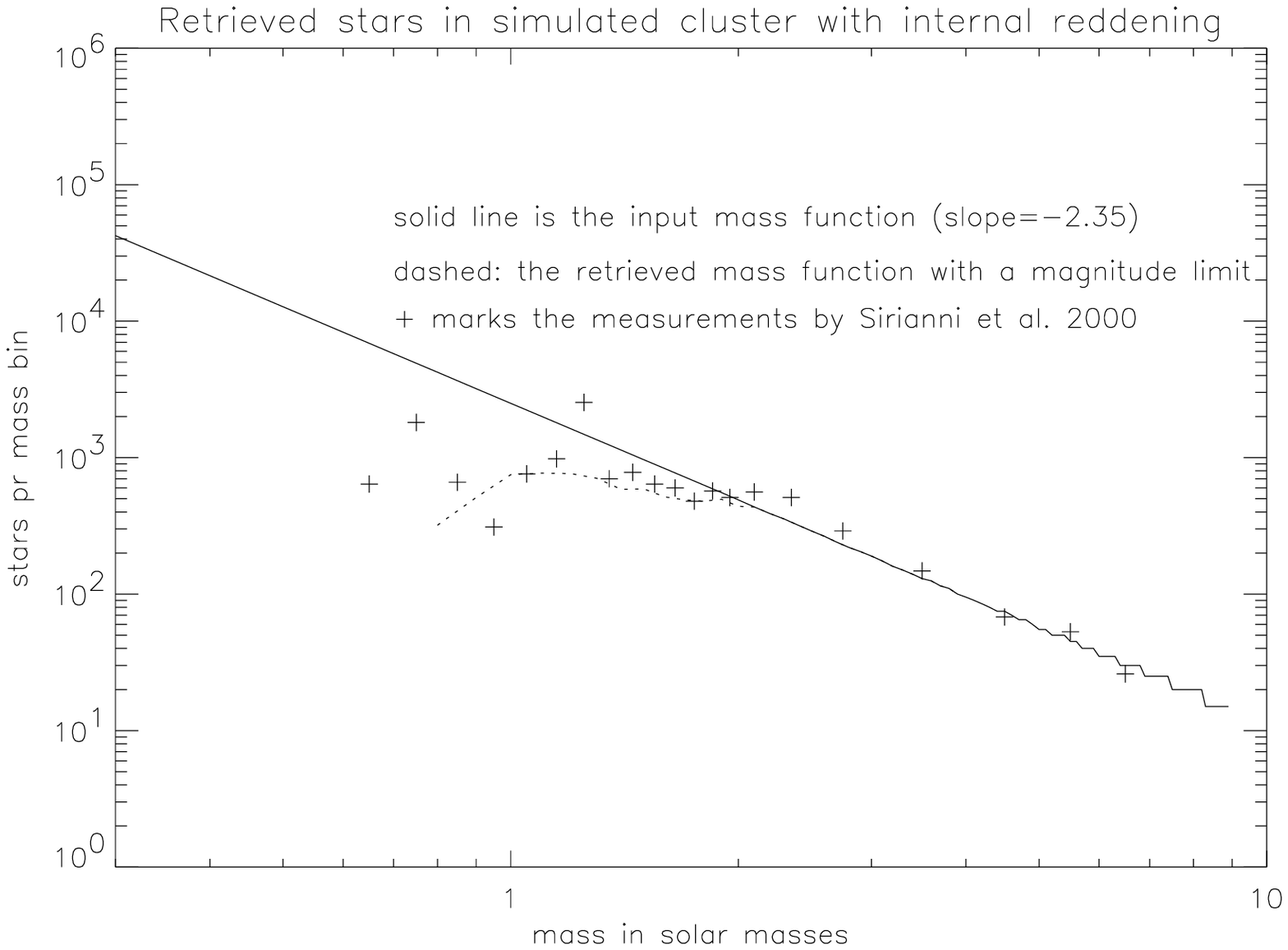,height=2in}}
\caption{The influence of differential extinction on the observed IMF
         slope.  For a constant detection limit (dashed line) variable
         reddening can cause an artificial ``flattening''.  Figure
         from \cite{and04a}.}
\label{sirianni}}
{\centerline{\psfig{file=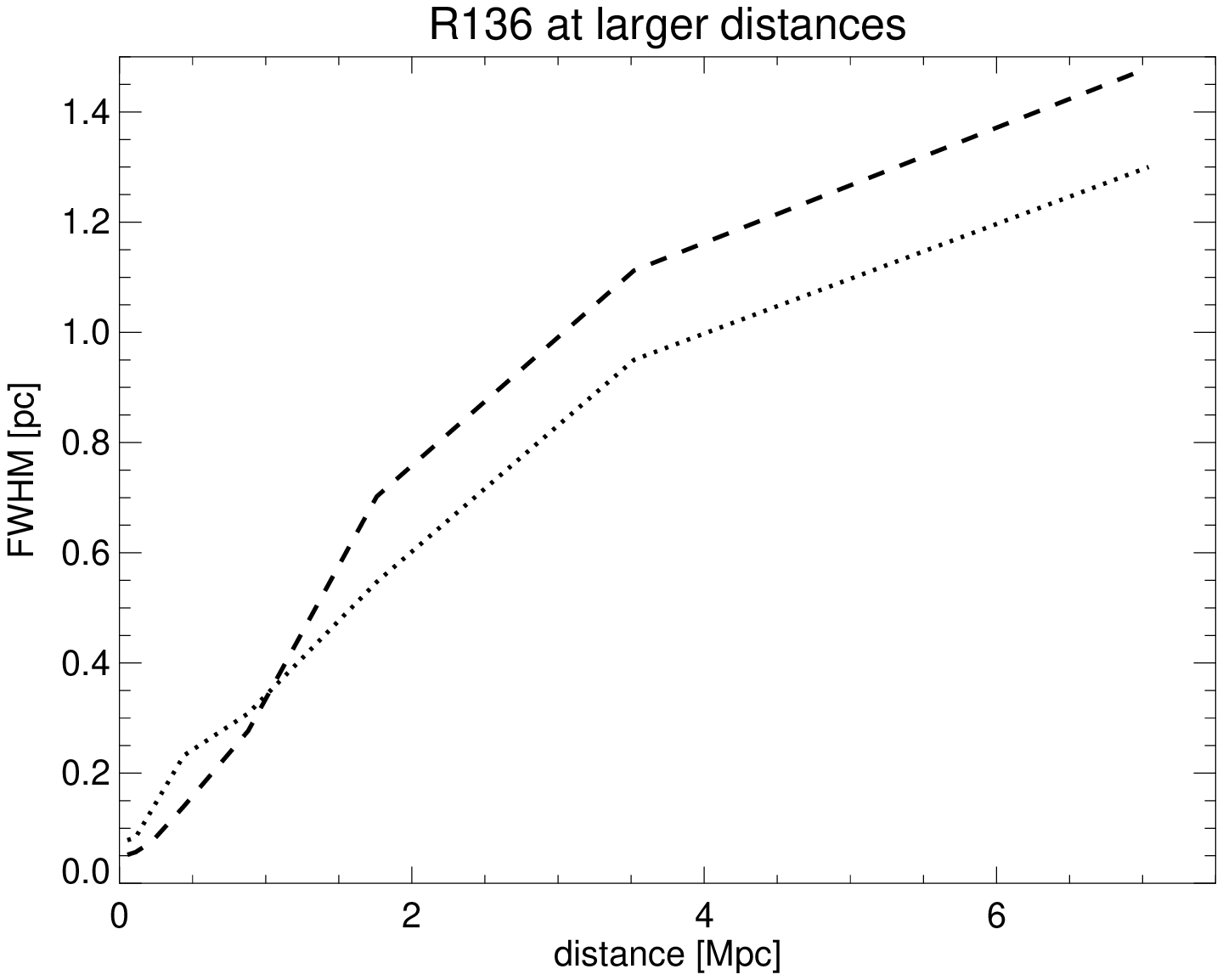,height=2in}}
\caption{The effect of spatial resolution on the derived half-light
         radius $r_{hl}$, measured for R136 locally, and remeasured at
         larger distances \cite{bra05}.}
\label{hubblerebin}}
\end{figure}

\cite{sir00} reported a ``definitive flattening below
$\approx 2 M_\odot$'' of the IMF slope at $r\ge 1.5$\,ps around R136.
This finding was based on optical data from HST/WFPC2 and the numbers
were corrected for incompleteness.  Unfortunately, the standard
Monte-Carlo technique for incompleteness correcton -- adding
artificial stars to the ``real'' image and computing their detection
probability as a function of distance from the cluster center and
source brightness -- corrects only for statistical (crowding,
blending) and instrumental (noise, flatfield) effects.  In contrast,
strong variations in the more localized {\sl differential} extinction
are caused either by the evolution of the PMS object or by rather
compact and patchy foreground extinction, and can vary by several
magnitudes at optical wavelengths.  These variations cannot be easily
corrected with Monte-Carlo techniques (see Figure~\ref{sirianni}).
Since the reddening distribution is very uncertain, the best
work-around are observations at near-IR wavelength where the
extinction is reduced by about an order of magitude.

\section{Method II:  cluster mass-to-light ratios}
Most starburst systems are located well beyond the Local Group and are
not resolvable into their individual stellar members (except for the
very brightest stars) with current techniques.  However, so-called
super star clusters (SSCs) may be the result of a recent starburst,
and their integrated properties can be observationally studied.

\subsection{The basic principle}
For a given set of assumptions (see below) the dynamical mass
$M_{dyn}$ of a cluster is related to its velocity dispersion $\sigma$
via
\begin{equation}
M_{dyn} = \eta \frac{\sigma^2 r_{hl}}{G}
\end{equation}
where $\eta \approx 10$.  The velocity dispersion $\sigma$ can be
determined from high resolution spectroscopy and the half-light radius
$r_{hl}$ from high resolution imaging.  Photometry will yield the
total luminosity $L$ and the mass-to-light ratio $M/L$ will provide an
estimate of the underlying IMF.  

Using this technique \cite{men02} found that the IMF in selected SSCs
in the Antennae galaxies is approximately consistent with Salpeter for
a mass range of $0.1 - 100 M_\odot$.  On the other hand, \cite{smi01}
found that in M82-F a lower mass cut-off of $2 - 3 M_\odot$ is
required for a MF with a Salpeter slope, i.e., {\sl M/L} is $5\times$
lower than normal.

\subsection{The problematic assumptions}
The approach described above depends on the validity of several
assumptions:

\noindent $\bullet$ {\sl The correct half-light radius $r_{hl}$.} Even
gravitationally-bound, young clusters often deviate from radially
symmetric, King-like cluster profile.  A perfect example is R136,
which has several bright cluster members at larger distance from the
center.  While those stars can be easily excluded from the light
profile at the distance of the LMC, they blend with the core at larger
distances (lower spatial sampling) and increase the measured $r_{hl}$.
This effect is illustrated in Figure~\ref{hubblerebin}.

\noindent $\bullet$ {\sl A gravitationally bound cluster.} While the
two most massive HII regions in the Local Group, 30\,Doradus in the
LMC and NGC\,604 in M\,33, have about the same luminosity they have
completely different morphological structures.  The core of 30\,Dor,
R136, is a massive compact cluster and presumably gravitationally
bound, while NGC\,604 consists of a several smaller, less luminous HII
regions, which are not gravitationally bound together.  However, at
the distance of e.g., the Antennae galaxies, where the scale of one
WFPC2 pixel corresponds to about 9 parsecs, 30\,Dor and NGC\,604 would
be indistinguishable.

\noindent $\bullet$ {\sl A constant M/L.} Massive stars appear to be
more concentrated toward the cluster center, see e.g., figure~16 in
\cite{bra96}.  We cannot distinguish here whether this is related to
the star formation process (i.e., massive stars form in the densest
regions), or the result of mass segregation.  At any rate, the most
massive stars dominate the light profile, and their concentration
around the cluster center produces a gradient in {\sl M/L}.

While the above method certainly allows for rough estimates of
dynamical cluster masses, the combined uncertainties make it very
difficult to provide sufficient evidence for anomalous IMFs in
galaxies beyond the Local Group.

\section{Method III:  mid-IR fine-structure line ratios}
Luminous starbursts are often significantly extincted and require
observations at longer wavelengths.  The mid-IR spectral range covered
by ISO-SWS and Spitzer-IRS provides a wealth of spectral diagnostic
features, although the spectra contain only spatially integrated
information.

\begin{figure}[ht]
\sidebyside
{\centerline{\psfig{file=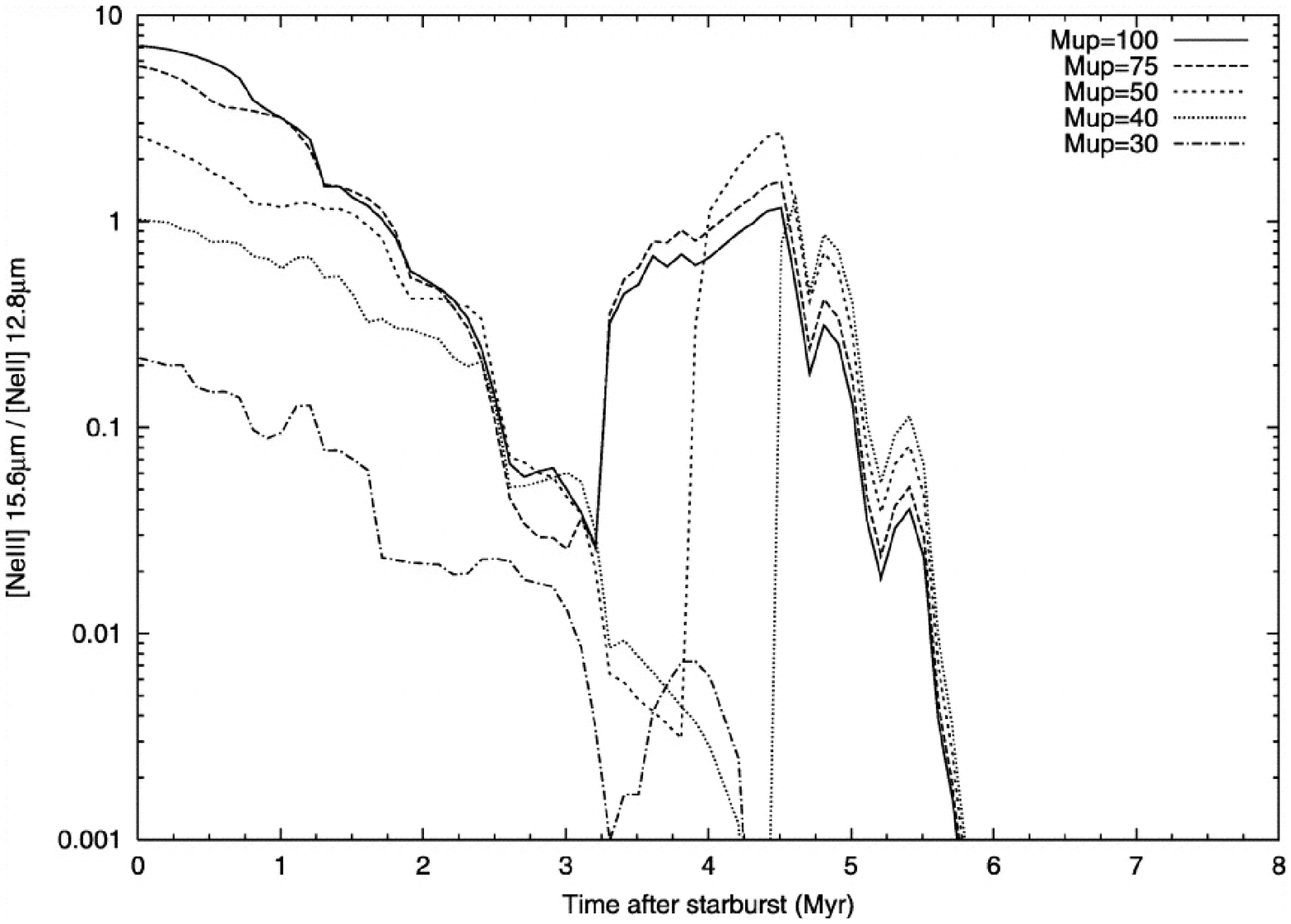,height=2in}}
\caption{Model ratios of the mid-IR [Ne\,III]/[Ne\,II] finestructure
         lines as a function of time and upper mass cut-offs.  Figure
         from \cite{rig04}.}
\label{rigby}}
{\centerline{\psfig{file=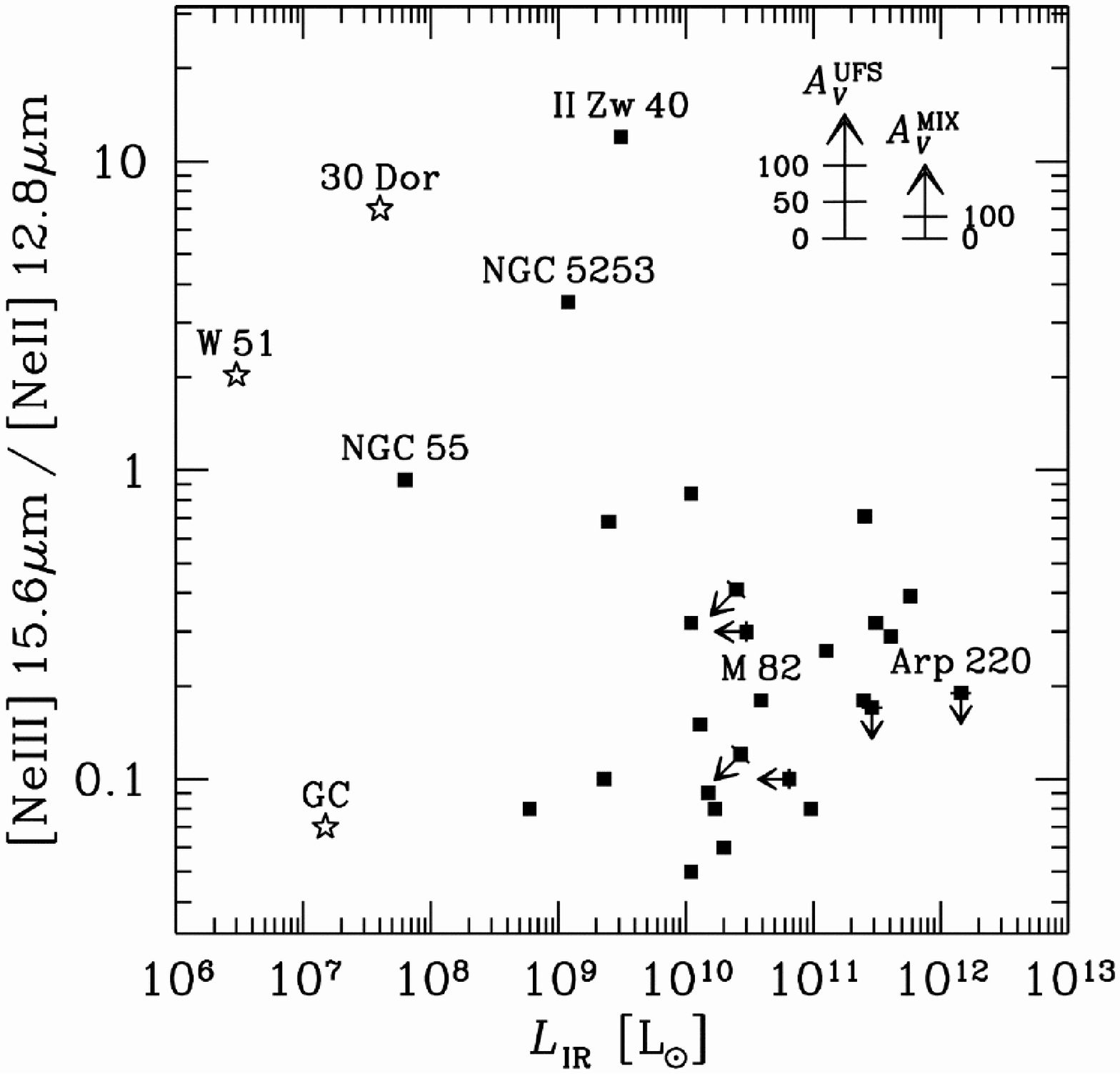,height=2in}}
\caption{Observed [Ne\,III]/[Ne\,II] ratios from ISO-SWS for a
         variety of Galactic and extragalactic sources.  The majority
         of luminous starburst galaxies fall below a ratio of one.
         Figure from \cite{tho00}.}
\label{thornley}}
\end{figure}

\subsection{The basic principle}
One of the most common diagnostics are the forbidden fine structure
lines [Ne\,III]$15.56\mu$m and [Ne\,II]$12.81\mu$m.  With an
ionization potential of 63.45\,eV, [Ne\,III] indicates the presence of
early-type O-stars, while [Ne\,II] with only 40.96\,eV can be excited
by less massive OB stars.  Thus, the ratio of [Ne\,III]/[Ne\,II] is
a very sensitive measure of the strength of the radiation field, which
depends on the presence of the most massive stars and on the age of
the starburst, as illustrated in Figure~\ref{rigby}.

\subsection{Problem: the model-observation discrepancy}
The observed line ratios (Figure~\ref{thornley}) differ significantly
from the values predicted by the models such as STARBURST99/CLOUDY
\cite{lei99}.  While there seems to be a better agreement for
low-metallicity galaxies and dwarf systems, the observed ratios of
more luminous starburst galaxies are often more than an order of
magnitude lower than the theoretical predictions.

Among the possible explanations being discussed is strong dependency
on metallicity, a general lack of the most massive stars in intense
bursts, or a larger age spread with complicated morphology of the
starburst region.  Higher extinction can be ruled out as it would even
increase the discrepancy (Figure~\ref{thornley}).  \cite{rig04}
proposed that very massive stars in dense and metal-rich starbursts
spend a significant fraction of their lifetime embedded in
ultra-compact HII regions that prevent the formation of classical HII
regions.  Hence, the finestructure lines expected from those embedded
massive stars cannot even form, and the spectra would be dominated by
the lower excitation [Ne\,II] line produced by the surrounding, less
massive stars.

In summary, mid-IR finestructure line diagnostics are an elegant and
intriguing method to infer the relative amount of massive stars in a
starburst.  However, given the complexity of a real starburst region
and the model uncertainties claims of anomalous IMFs, just based on
line ratios, need to be taken with care.  This is clearly an area
where significant improvements can be expected from Spitzer-IRS
observations.



\end{document}